\documentclass[twocolumn,preprintnumbers,amsmath,amssymb,showkeys]{revtex4}
\usepackage{url}
\usepackage{amsmath}
\usepackage{amssymb}
\usepackage{graphicx}

\begin{document}

\title{Statistical Invisibility of a Physical Attack on QRNGs After Randomness Extraction}

\author{Yi-Fan Chen$^{1,*}$, Dong Wang$^{4,}$\footnote{The authors Yi-Fan Chen and Dong Wang contributed equally to this work.}, Yi-Bo Zhao$^{2,3,4,}$\footnote{Corresponding author: zhaoyb@hizju.org}, Liang Cheng$^{1,}$\footnote{Corresponding author: chengliang@iscas.ac.cn}, Yi Zhang$^5$, and Yang Zhang$^{1,6}$}
\affiliation{%
$^1$Institute of Software, Chinese Academy of Sciences, Beijing, China\\
$^2$Huzhou Institute of Zhejiang University, Huzhou 313000, China \\
$^3$College of Control Science and Engineering, Zhejiang University, Hangzhou 310027, China \\
$^4$Beijing GGQuanta Co. Ltd., Beijing, China \\
$^5$Institute of Cryptography and Cyberspace Security (Huangpu), Guangzhou, China\\
$^6$Zhongguancun Laboratory, Beijing, China\\
}%

\date{\today}

\begin{abstract}

Current prevailing designs of quantum random number generators (QRNGs) designs typically employ post-processing techniques to distill raw random data, followed by statistical verification with suites like NIST SP 800-22. This paper demonstrates that this widely adopted practice harbors a critical flaw. We show that the powerful extraction process can create a false sense of security by perfectly concealing physical-layer attacks, rendering the subsequent statistical tests blind to a compromised entropy source. We substantiate this claim across two major QRNG architectures. Experimentally, we severely compromise an QRNG based on amplified spontaneous emission (ASE) with a power supply ripple attack. While the resulting raw data catastrophically fails NIST tests, a standard Toeplitz extraction transforms it into a final sequence that passes flawlessly. This outcome highlights a profound danger: since the validation process is insensitive to the quality of the raw data, it implies that even a fully predictable input could be processed to produce a certified, yet completely insecure, random sequence. Our theoretical analysis confirms this vulnerability extends to phase-noise-based QRNGs, suggesting a need for security validation to evolve beyond statistical analysis of the final output and consider the entire generation process.

\end{abstract}

\maketitle


\section{Introduction}

The integrity of modern cryptographic systems, from secure communication protocols to digital signatures, is predicated upon a continuous supply of unpredictable random numbers~\cite{rfc4086, 7782774}. Traditionally, pseudo-random number generators (PRNGs) have been widely adopted because of their efficiency and ease of implementation~\cite{rukhinStatisticalTestSuite2010}. PRNGs generate sequences based on deterministic algorithms and initial seeds, meaning that their randomness fundamentally depends on the complexity of the algorithms and the secrecy of the seeds. This type of randomness sources are theoretically predictable: If the algorithms are compromised or the seeds exposed, the security of the entire system is at risk~\cite{kelsey1998cryptanalytic, heninger2012mining}. This inherent predictability makes them untenable for applications demanding information-theoretic security, such as quantum key distribution (QKD)~\cite{RevModPhys.81.1301}.

The emergence of QRNGs offers a fundamental solution to this problem. Unlike PRNGs, QRNGs harness the intrinsic randomness of quantum mechanical processes, such as single-photon detection~\cite{schranz2024stochastic,amri2020single}, multi-photon detection~\cite{aungskunsiri2023quantum}, phase fluctuations~\cite{lei20208}, vacuum state fluctuations~\cite{gabriel2010generator}, or ASE noise\cite{guo202140}, as entropy sources, ensuring that the generated sequences are truly unpredictable and genuinely random. This physical, rather than algorithmic, origin of randomness makes QRNGs an ideal choice for information-theoretically secure random number generation. Among these, schemes based on ASE noise of the SLEDs~\cite{ase} and those based on laser phase noise~\cite{xu2012ultrafast} have attracted tremendous attention due to their potential for high-speed and robust integration. 

Despite the inherent randomness of quantum processes, the practical security of QRNGs depends heavily on the integrity and robustness of their system implementation against environmental variations and interference~\cite{herrero2017quantum}. Studies have shown that interference with entropy sources, even caused by normal changes in environmental conditions, can significantly affect the statistical properties of the entropy source, such as the output optical power distribution and min-entropy~\cite{bayon2012contactless}. More concerningly, this sensitivity can indeed be weaponized~\cite{markettos2009frequency}: By carefully designing manipulations at the physical layer, such as injecting specific periodic disturbance signals into the drive circuit, an adversary can compromise the inherent randomness of the entropy source or even implant predictable patterns, thereby threatening the security of the entire QRNG system.

To ensure the quality of the final output, practical QRNGs typically employ key procedures for purification and validation. The raw data from the physical entropy source is first refined using a randomness extractor—a post-processing algorithm designed to distill entropy and remove inherent biases. Subsequently, the resulting sequence is evaluated against a suite of statistical tests to check for any remaining non-random patterns. In this context, the combination of Toeplitz hashing for extraction and the NIST SP 800-22 suite~\cite{rukhinStatisticalTestSuite2010} for validation has become a common and trusted approach in QRNG implementations~\cite{maPostprocessingQuantumRandomnumber2013, imran2020quantum, bai202118, alvarez2020monolithic, imran2021chip}.

However, our work reveals a critical flaw in this widely used methodology. We find that the post-processing step designed to remove statistical imperfections can also erase the tell-tale signs of a physical-layer attack. But existing standards for evaluating random number generation techniques focus primarily on analyzing the statistical properties of the generated random numbers, including uniformity, independence, and complexity. Rarely do they consider the integrity of the entropy source or the generation process. As results, these standards are effective in detecting algorithmic flaws or statistical biases during random number generation, but share the same limitation in distinguishing whether or not a statistically compliant number sequence originates from a degraded or even manipulated entropy source~\cite{herrero2017quantum}. This means NIST SP 800-22 suite is effectively blind to the initial state of the entropy source, creating a potential for a false sense of security where a compromised device is certified as sound.

Our work focuses on two prevalent QRNG architectures: those based on ASE noise and laser phase noise, both of which are widely adopted in commercial products worldwide. First, we design and validate attacks against these QRNGs. We demonstrate the attack's effectiveness by directly modulating the drive current of the ASE-based QRNG - a technique we term the power supply ripple attack (detailed further in section \ref{sec:attack principle}). Empirical results show this attack significantly compromises the raw data from it. We then generalize these findings on the theoretical model for the phase-noise-based QRNG, which is also vulnerable to the similar physical mechanism. This theoretical confirmation is sufficient, as our paper's focus is not the attack itself, but its utility in testing post-processing. We use the compromised data to demonstrate that a randomness extractor, by design, can restore its statistical uniformity to the point that it passes all validation tests, effectively masking the total collapse of the physical entropy source. Specifically, after the compromised data is processed by a Toeplitz hashing extractor, the final sequence successfully passes the entire NIST SP 800-22 suite. The use of such a compromised QRNG has severe security implications: cryptographic keys, nonces, or other secret tokens generated from it become predictable to attackers, potentially leading to complete system compromise.

The rest of this paper is organized as follows. Section \ref{sec:background} provides the necessary background on the two targeted QRNG architectures, the rationale for the power supply ripple attack, and the standard validation pipeline comprising Toeplitz hashing and NIST tests. Sections \ref{sec:ase_attack} and \ref{sec:phase_attack_model} then establish the efficacy of the attack, first through an experimental demonstration on the ASE-based QRNG and then via a theoretical model for the phase-noise-based scheme. Section \ref{sec:limits_validation} presents our central finding: the post-processed output from the compromised source passes the entire NIST test suite. Finally, Section \ref{sec:conclusion} concludes the paper and discusses the security implications.

\section{Background}
\label{sec:background}

\subsection{Target Quantum Random Number Generation Schemes}
\label{sec:target-QRNG}

Our work focuses on two prominent types of QRNGs, based on ASE noise and laser phase noise, chosen for their high-speed potential and prevalence in both academic research and commercial products.

Within modern fiber optic telecommunications, ASE represents a fundamental and pervasive source of noise whose statistical characteristics have been extensively studied. And ASE-based QRNGs have attracted tremendous attention due to their advantages in achieving high generation rates. Williams et al. first reported on the work of generating random bits using filtered ASE produced in a fiber amplifier and achieved random number generation of 12.5 Gb/s in 2010~\cite{williams2010fast}. In 2014, Li et al. achieved robust real-time random bit sequences at rates up to 2.5 Gbit/s using filtered ASE from the the super-luminescent diode (SLD)~\cite{li2014random}. In 2017, Wei et al. presented a compact and high-speed QRNG with a real-time generation rate of 1.2 Gbps based on measuring the ASE noise of the SLED~\cite{wei2017compact}. More recently, Guo et al. developed a fast QRNG based on optically sampled ASE that continuously generates a 40 Gb/s random bit sequence~\cite{guo202140}. The experimental platform in this work is modeled after the QRNG proposed by Wei et al. We selected this architecture as our target because its combination of real-time performance, high speed, and compact size makes it a prime example of a system with significant potential for practical and widespread deployment.

Another popular high-speed scheme utilizes the phase noise of a laser. Phase fluctuations (or noise) can be attributed to spontaneous emission, which can be used as a quantum random source. To access this randomness, an interferometer is necessary to convert these unmeasurable phase fluctuations into detectable intensity variations. However, this quantum signal is typically superimposed with classical noise, so maximizing the quantum-to-classical noise contribution is critical for reliable detection. The work by Xu et al. in 2012~\cite{xu2012ultrafast} was a highly influential demonstration, successfully generating random bits at over 6 Gb/s by harnessing quantum phase fluctuations and highlighting the immense potential of this approach. Following this, many other researchers have developed QRNGs based on phase noise, achieving higher generation rates and real-time operation~\cite{abellan2014ultra, nie2015generation, lei20208}. The theoretical security analysis for phase-noise-based QRNGs presented in this paper is based on the scheme established in the work of Xu et al. because of the simplicity and high speed of their experimental setup.

\subsection{Power Supply Ripple Attack}
\label{sec:attack principle}

While ideal QRNGs are perfectly secure, their practical implementations are physical systems susceptible to environmental influences and side-channel attacks. The security of the final random numbers depends on the physical integrity of the entropy source. SLED sources used in both ASE-based and phase-noise-based QRNGs are known to be sensitive to their operating parameters, particularly the electrical drive current and temperature. Li et al.~\cite{liExperimentalStudySecurity2021} experimentally characterized how variations in drive current and temperature impact the statistical properties and min-entropy of an SLED-based QRNG. Their work rigorously demonstrated that changes in these operating conditions degrade the quality of the raw randomness.

Our work takes this concept one step further. Instead of passively observing the effects of environmental drift, we treat this dependency as an active attack vector. This direction was notably advanced by Beijing GGQuanta Co. Ltd., which in September 2024 first proposed the impact of power supply ripple attacks on quantum random number generators\cite{gg1, gg2}. A power supply ripple attack leverages this sensitivity by superimposing a controlled, periodic signal onto the device's DC drive current. The injected ripple can remain hidden within normal supply fluctuations. By doing so, an adversary can effectively influence the physical entropy source and imprint a predictable pattern onto the raw data, gaining fine-grained control over the QRNG's output. Our goal is not to claim novelty in the attack method itself, but to use it as a simple and reproducible tool to controllably compromise the raw data. This approach establishes the necessary foundation to demonstrate the critical flaw in the standard validation pipeline: powerful post-processing may erase the evidence of a physical attack, allowing a compromised QRNG output to pass all standard statistical tests.

\subsection{Toeplitz Hash Extractor and NIST Statistical Test Suite}

The inherent quantum randomness in the output of a QRNG is typically mixed with classical noise. To counteract potential attacks exploiting classical or quantum side information, randomness extraction (or post-processing) is essential. This process distills genuine randomness from the mixture of quantum and classical noise. There are many post-processing methods, including Von Neumann's post-processing~\cite{von195113}, XOR operation\cite{qi2010high}, Toeplitz-Hash extractor~\cite{lin2024seed}, and so on. Among these, the Toeplitz-hash extractor is particularly valuable due to its information-theoretic security and efficiency. Toeplitz hash extractor has a significantly faster speed thanks to its relatively simple structure. It has been used in privacy amplification and adapted for QRNGs to address similar challenges. In practice, the Toeplitz extractor employs a constructed Toeplitz matrix to convert an arbitrary input distribution into a uniform distributed output.

The de facto standard for validating the statistical quality of random number generators is the NIST SP 800-22 test suite~\cite{rukhinStatisticalTestSuite2010}. It comprises a battery of 15 statistical tests, each designed to detect specific non-random characteristics that might be present in a sequence. The most basic test is the Frequency test. It is first performed and a failure in this will lead to the failure of subsequent tests. For a sequence to be considered random, it must pass all applicable tests. Some key tests in the suite include:

\begin{itemize}
    \item \textbf{Frequency Test:} Checks if the proportion of zeros and ones is close to 0.5, testing for basic bias.
    \item \textbf{Runs Test:} Checks for the proper oscillation between runs of zeros and ones. Too few or too many runs suggest a lack of independence.
    \item \textbf{Fast Fourier Transform (FFT) Test:} Detects periodic features in the sequence, which would indicate underlying regularity.
    \item \textbf{Linear Complexity Test:} Assesses whether the sequence can be easily generated by a linear feedback shift register, a measure of its unpredictability.
    \item \textbf{Approximate Entropy Test:} Measures the frequency of all possible overlapping blocks of adjacent bits, checking for uniformity and regularity over different block sizes.
\end{itemize}

\section{Attack on ASE-based QRNG}
\label{sec:ase_attack}

\subsection{Experimental Setup}
\label{sec:experiment-setup}

\begin{figure}[htbp]
    \centering
    \includegraphics[width=\linewidth]{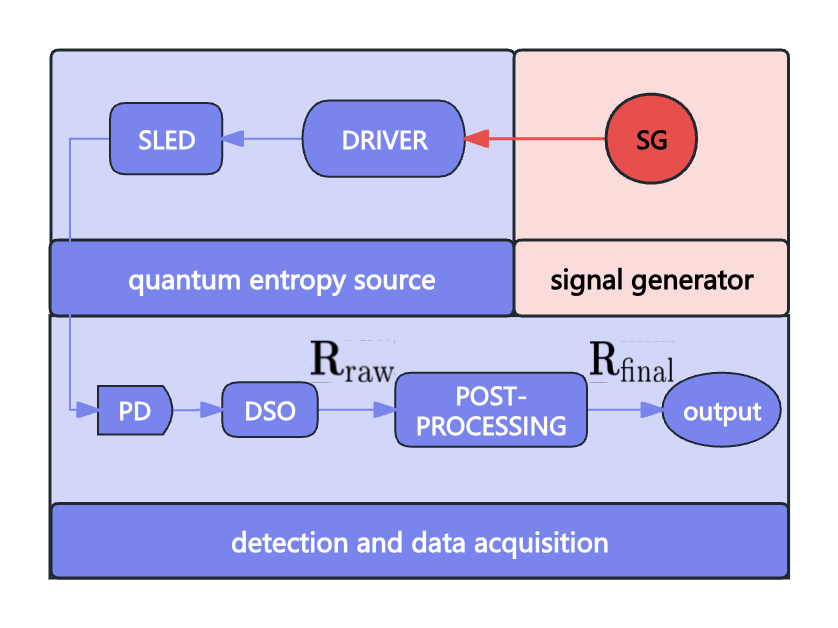}
    \caption{General architecture of the ASE-based QRNG under study. The entropy source is the ASE noise from the SLED. The attack is injected into the drive current via a signal generator.}
    \label{fig:sled}
\end{figure}

The experimental setup for our attack is modeled after the ASE-based QRNG architecture from Wei et al., which was selected as our target in Section\ref{sec:target-QRNG}. A schematic of this setup is depicted in Fig.~\ref{fig:sled}.

In this scheme, the SLED's ASE noise serves as the physical entropy source. As shown in the figure, the optical signal is converted into an electrical signal by a photodiode (PD), and then digitized by a high-resolution Digital Storage Oscilloscope (DSO). This process yields the unprocessed bit sequence, denoted as $\mathbf{R}_{\mathrm{raw}}$. The final random data, $\mathbf{R}_{\mathrm{final}}$, is obtained after applying a post-processing extractor to $\mathbf{R}_{\mathrm{raw}}$.

Our attack model directly exploits the SLED's known sensitivity to its drive current. We inject a controlled, periodic signal from an external signal generator onto the DC drive current. The objective is to induce intensity fluctuations that dominate the intrinsic quantum noise. By doing so, we aim to imprint a predictable and classical pattern onto the raw data $\mathbf{R}_{\mathrm{raw}}$, thereby completely compromising the integrity of the physical entropy source. This controlled compromise provides the groundwork for our subsequent investigation: testing whether the post-processing can mask such a fundamental failure.

\subsection{Implementation and Efficacy Validation}
\label{sec:experiment}

We then implemented the attack on our experimental platform, which follows the architecture described in Sec~\ref{sec:experiment-setup}. The SLED was operated under carefully selected conditions to ensure stable and optimal performance: a constant temperature of $25 \pm 0.1^\circ$C and a drive current of 40~mA, which places the device in a regime of high entropy output. The attack signal was chosen to be a 50~kHz square wave. The square wave provides sharp, rapid transitions that induce a strong, predictable pattern in the SLED's output, while the 50~kHz frequency was selected to be well within the system's overall bandwidth. The resulting optical signal was detected and then digitized by a Pico 3206D DSO at its native sampling rate of 9.62MS/s. This rate significantly oversamples the 50kHz attack signal, ensuring the compromised waveform is captured with high fidelity and without aliasing. Finally, the raw samples were binarized using a zero-crossing threshold, a standard method for converting an AC-coupled noise signal into a bitstream.

To quantitatively evaluate the attacker's control over the entropy source, we calculated the Pearson correlation coefficient $\rho$ between the raw binary sequence from the attacked QRNG, $\mathbf{R}_{\mathrm{raw}}^{\mathrm{attack}}$, and a reference sequence, $\mathbf{R}_{\mathrm{raw}}^{\mathrm{square}}$, derived by applying the identical binarization process to the injected square-wave signal itself. Both sequences underwent DC removal and binarization. We swept the square-wave peak-to-peak voltage ($V_{\mathrm{pp}}$) from $200~\mathrm{mV}$ to $1000~\mathrm{mV}$ in increments of $100~\mathrm{mV}$. At each voltage level, three acquisitions of $6,\!153,\!984$-bit sequence pairs were conducted independently. The correlation coefficient $\rho$ was calculated for each pair and then averaged across the three runs. 

\begin{figure}[htbp]
    \centering
    \includegraphics[width=\linewidth]{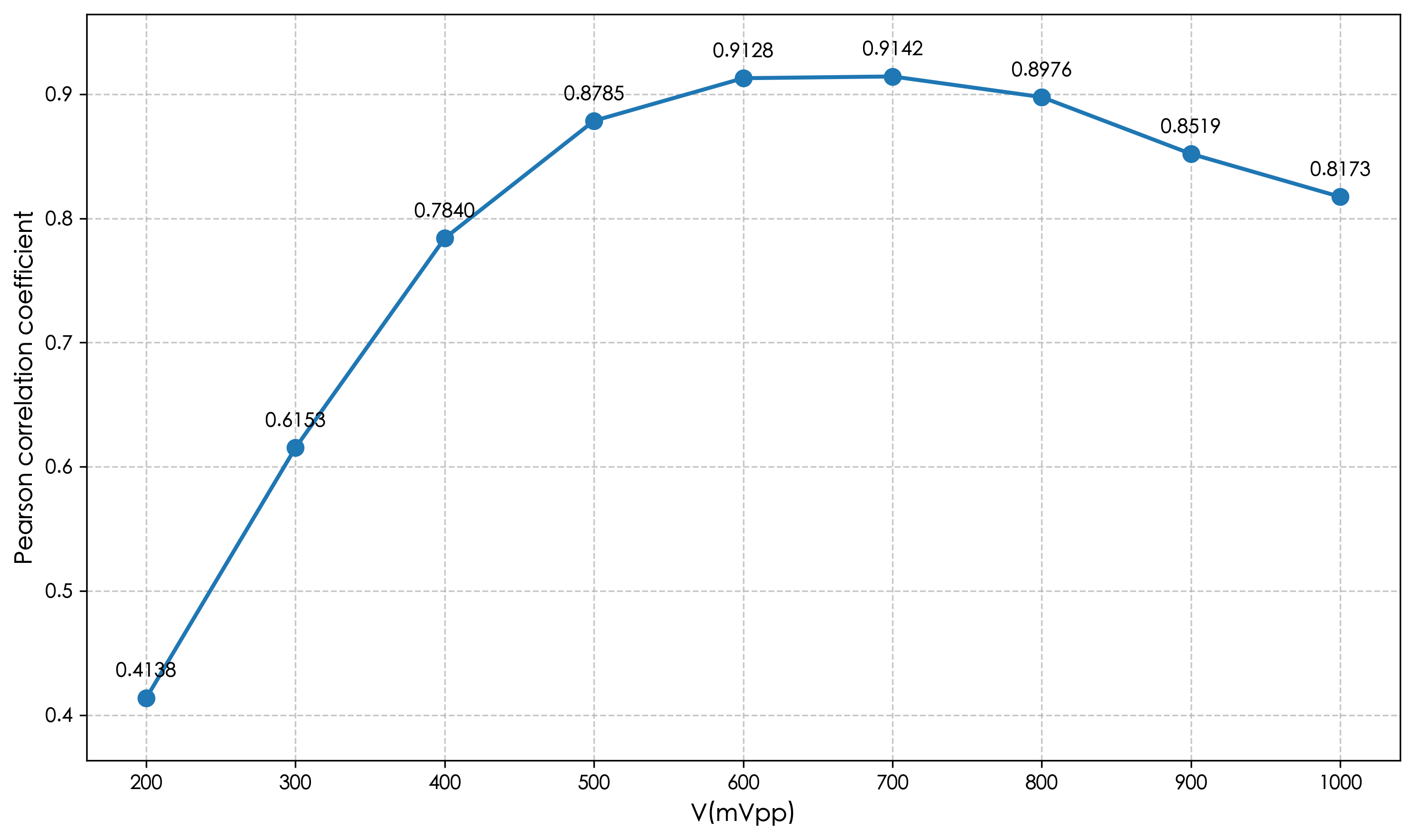}
    \caption{Pearson correlation coefficient as a function of the attack signal's voltage amplitude. The correlation peaks at 0.9142 for an attack voltage of 700 mV, indicating a near-total compromise of the entropy source at the physical level.}
    \label{fig:pearson}
\end{figure}

As shown in Figure~\ref{fig:pearson}, the mean correlation coefficient increases rapidly from 0.41 when $V_{pp}$ is set to 200~mV to a maximum of 0.9142 when $V_{pp}$ is 700~mV. Increasing $V_{pp}$ beyond 800~mVpp causes the correlation coefficient to slightly decreases, which can be attributed to the increased jitters in the laser output caused by excessive ripple. Based on this scan, we selected 700~mVpp as the optimal injection parameter for subsequent power supply ripple attack experiments to maximize disturbance to the entropy source.

To further illustrate the impact of this compromise, we performed a comparative statistical analysis using the NIST SP 800-22 suite. We tested the raw data from the baseline (unattacked) system, $\mathbf{R}_{\mathrm{raw}}^{\mathrm{base}}$, against the data from the optimally attacked system, $\mathbf{R}_{\mathrm{raw}}^{\mathrm{attack}}$. In our tests, each raw sequence of 6,153,984 bits was divided into 10 bit streams for the NIST tests. According to the test requirements, at least eight out of 10 bit streams must pass each test item for the result to be considered successful. The results are shown in Table~\ref{tab:pretest}.

\begin{table}[htbp]
    \centering
    \begin{tabular}{lcc}
        \hline
        \textbf{Test Category} & \textbf{Baseline} & \textbf{Ripple-Attack} \\
        \hline
        Frequency & Passed & Failed \\
        BlockFrequency & Passed & Failed \\
        CumulativeSums & Failed & Failed \\
        Runs & Failed & Failed \\
        LongestRun & Passed & Failed \\
        Rank & Passed & Failed \\
        FFT & Passed & Failed \\
        NonOverlappingTemplate & Passed & Failed \\
        OverlappingTemplate & Passed & Failed \\
        Universal & Passed & Failed \\
        ApproximateEntropy & Passed & Failed \\
        RandomExcursions & - & - \\
        RandomExcursionsVariant & - & - \\
        Serial & Passed & Failed \\
        LinearComplexity & Passed & Failed \\
        \hline
        \textbf{Overall Result} & \textbf{Failure} & \textbf{Failure} \\
        \hline
    \end{tabular}
    \caption{NIST SP 800-22 test results for raw random sequences. The attacked data fails all tests, confirming the attack's effectiveness at the raw data level. The Random Excursions tests are not applicable (-) as the raw data did not meet the prerequisite number of cycles.}
    \label{tab:pretest}
\end{table}

As expected, the baseline raw data, $\mathbf{R}_{\mathrm{raw}}^{\mathrm{base}}$, fails several tests due to inherent biases from the physical hardware, though it still passes a majority of the more complex statistical checks. In stark contrast, the compromised sequence, $\mathbf{R}_{\mathrm{raw}}^{\mathrm{attack}}$, fails every applicable test. This catastrophic failure is further evidenced by the frequent "igamc: UNDERFLOW" errors reported by the NIST suite, indicating a complete breakdown of statistical randomness. 

Taken together, the near-unity correlation and the total failure of NIST tests provide unequivocal evidence that our attack has successfully and completely compromised the raw random data from QRNG. The raw data is, for all practical purposes, predictable. This establishes the critical baseline for the central question of our study: whether the post-processing is so effective at restoring statistical randomness that it allows the final data to pass all validation tests, thereby concealing the compromise of the physical source.

\section{Theoretical Analysis of an Equivalent Attack on Phase-Noise QRNGs}
\label{sec:phase_attack_model}

\begin{figure}[htbp]
    \centering
    \includegraphics[width=\linewidth]{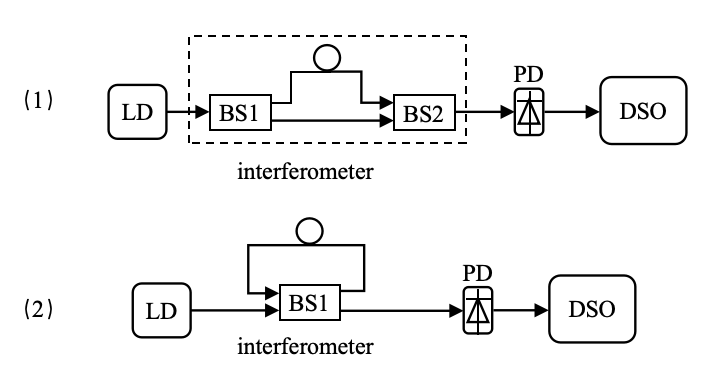}
    \caption{Two typical phase-noise-based QRNG schemes.}
    \label{fig:phase}
\end{figure}

To demonstrate the broader applicability of our attack methodology, we now extend our analysis to the second target QRNG type introduced in Section~\ref{sec:target-QRNG}: systems based on laser phase noise. These devices convert the random phase fluctuations of a laser into measurable intensity variations using a delayed self-interferometer, as shown in Fig.~\ref{fig:phase}. There are two typical interferometer structures (with arm delay $T$): one is an unbalanced Mach-Zehnder interferometer (MZI) constructed with two beam splitters (BS) and a fiber delay line; the other is a delay-line interferometer with a single beam splitter and a fiber delay. The function of the interferometer is to interfere the optical signal at time $t$ with that at time $t-T$. Suppose the laser is modulated by a periodic signal with period $T$. In this case, the intensities of the two interfering signals can be controlled by the modulation, allowing the output interference intensity to be manipulated.

Taking the first structure as an example, the output optical field of the laser can be defined as:
\begin{equation}
E_{\text{in}}(t) = A(t) e^{i[\omega t + \theta(t)]}
\end{equation}
where $A(t)$ and $\theta(t)$ are respectively the amplitude and phase of the optical signal at time $t$, and $\omega$ is the angular frequency.

After passing through the interferometer, the output field is:
\begin{equation}
\begin{split}
E_{\text{out}}(t) &= \frac{1}{2}\left[ E_0(t - \tau) + E_0(t) \right] \\
&= \frac{1}{2}\left\{ A(t - \tau) e^{i[ \omega(t - \tau) + \theta(t - \tau) ]} + A(t) e^{i[ \omega t + \theta(t) ]} \right\}
\end{split}
\end{equation}
The corresponding intensity is:
\begin{align}
I(t) &= \frac{1}{4}\Bigl[ A^2(t-\tau) + A^2(t) \notag \\
&\quad + 2A(t-\tau)A(t)\cos(\omega\tau + \Delta\theta(t)) \Bigr]
\label{eq:intensity}
\end{align}
where $\Delta\theta(t) = \theta(t) - \theta(t-\tau)$ is the phase difference between the two optical paths.

For a normally operating QRNG, the amplitude $A(t)$ can be considered as constant. That is, $A(t) = A$, so the intensity simplifies to $P = A^2$. The interference output intensity $I(t)$ thus becomes:
\begin{equation}
I(t) = \frac{1}{2} P \left[ 1 + \cos( \omega \tau + \Delta\theta(t) ) \right]
\end{equation}

After filtering out the DC component of the electrical signal from the photodiode (PD) and setting a constant phase bias of $2m\pi + \pi/2$ between the two arms, the resulting electrical signal is:
\begin{equation}
V(t) = k P \sin( \Delta\theta(t) ) \approx k P \Delta\theta(t)
\end{equation}
where $k$ is a constant that includes the PD detection efficiency and the amplifier gain. Since $\Delta\theta(t)$ is sufficiently small, $\sin(\Delta\theta(t)) \approx \Delta\theta(t)$. Thus, the PD output is proportional to the fluctuation of the random phase, which can be used for the generation of random numbers.

When the QRNG is subjected to the aforementioned attack, suppose $A(t)$ fluctuates around $A_0$ by an amount $\alpha$, the interference output intensity becomes:
\begin{equation}
\begin{split}
I(t) &= \frac{1}{4}\biggl[ (A_0 + \alpha)^2 + (A_0 - \alpha)^2 \\
     &\quad + 2(A_0^2 - \alpha^2)\cos(\omega\tau + \Delta\theta(t)) \biggr] \\
&= \frac{1}{2}\left[ A_0^2 + \alpha^2 + (A_0^2 - \alpha^2)\cos(\omega\tau + \Delta\theta(t)) \right] \\
&\approx \frac{1}{2}\left[ A_0^2 + \alpha^2 + (A_0^2 - \alpha^2)\Delta\theta(t) \right]
\end{split}
\end{equation}
Assuming $\alpha$ varies with time, i.e. $\alpha = \alpha(t)$ and $\Delta\theta(t) \ll 1$, after filtering the DC component, the electrical signal is:
\begin{equation}
\begin{split}
V(t) &= k\left[ \alpha^2(t)(1 - \Delta\theta(t)) + P\Delta\theta(t) \right] \\
&\approx k\left[ \alpha^2(t) + P\Delta\theta(t) \right]
\end{split}
\end{equation}

Therefore, by precisely controlling the waveform of $\alpha(t)$, an adversary can manipulate the sign of the output voltage: when $\frac{\alpha^2(t)}{P} > -\Delta\theta(t)$, the output is positive; when $\frac{\alpha^2(t)}{P} < -\Delta\theta(t)$, the output is negative. In this way, the attacker can effectively overwhelm the quantum randomness and dictate the binarized output of the QRNG. The attack scheme is theoretically equivalent to the power supply ripple attack on ASE-based QRNGs discussed in Section~\ref{sec:ase_attack}. In both scenarios, a carefully chosen periodic signal can be used to imprint a predictable classical pattern onto the physical entropy source, completely compromising its integrity before any post-processing is applied.

\section{The Limits of Statistical Testing}
\label{sec:limits_validation}

Randomness extraction is a critical component of any practical QRNG. Its purpose is to take a raw, potentially biased data stream and produce a final sequence that is statistically indistinguishable from a perfect uniform distribution. Among the most powerful and widely adopted techniques for this task is Toeplitz hashing. Based on the principles of universal hashing and underpinned by the Leftover Hash Lemma (LHL), a Toeplitz extractor can effectively suppress statistical defects and eliminate predictability from its input, provided the input retains sufficient min entropy.

This theoretical power, however, leads us to a critical security question: when a power supply ripple attack has successfully imprinted a predictable, non-random pattern onto the output raw random data, is the extractor's inherent ability to produce a statistically uniform output so effective that it conceals the complete failure of the compromised raw data stream from standard validation tests?

To answer this, we applied a Toeplitz hashing extractor to the two raw data streams generated in Section~\ref{sec:experiment}: the baseline data ($\mathbf{R}_{\mathrm{raw}}^{\mathrm{base}}$) and the compromised data ($\mathbf{R}_{\mathrm{raw}}^{\mathrm{attack}}$). We configured a $1024 \times 2048$ Toeplitz matrix, which compresses a 2048-bit raw input block into a 1024-bit final output (a compression ratio of 0.5). This configuration is not arbitrary; it is designed to meet stringent security requirements, satisfying the LHL with a security parameter $\epsilon < 2^{-20}$ for any input with sufficient min-entropy~\cite{lei84GbpsRealtime2020,symulRealTimeDemonstration2011}. The resulting final sequences, $\mathbf{R}_{\mathrm{final}}^{\mathrm{base}}$ and $\mathbf{R}_{\mathrm{final}}^{\mathrm{attack}}$, were then subjected to the full NIST SP 800-22 test suite.

\begin{table}[htbp]
    \centering
    \begin{tabular}{lcc}
        \hline
        \textbf{Test Category} & \textbf{Baseline} & \textbf{Ripple-Attack} \\
        \hline
        Frequency & Passed & Passed \\
        BlockFrequency & Passed & Passed \\
        CumulativeSums & Passed & Passed \\
        Runs & Passed & Passed \\
        LongestRun & Passed & Passed \\
        Rank & Passed & Passed \\
        FFT & Passed & Passed \\
        NonOverlappingTemplate & Passed & Passed \\
        OverlappingTemplate & Passed & Passed \\
        Universal & Passed & Passed \\
        ApproximateEntropy & Passed & Passed \\
        RandomExcursions & Passed & Passed \\
        RandomExcursionsVariant & Passed & Passed \\
        Serial & Passed & Passed \\
        LinearComplexity & Passed & Passed \\
        \hline
        \textbf{Result} & \textbf{Success} & \textbf{Success} \\
        \hline
    \end{tabular}
    \caption{NIST test results for the final random sequences after post-processing under two acquisition scenarios.}
    \label{tab:posttest}
\end{table}

The results indicate that, regardless of whether the data came from the baseline group or the ripple attack group, the sequences generated after Toeplitz hashing passed all basic and advanced statistical tests. Although the power supply ripple attack implanted predictable bias patterns in the raw data, Toeplitz hashing not only corrected the statistical bias of the output bits, but also "sanitized" and concealed all traces of physical-layer manipulation, rendering the NIST statistical tests completely ineffective. This demonstrates that, when information-theoretically secure post-processing is applied, "over-sanitization" may create a false sense of security, and relying solely on statistical tests of the output sequence is insufficient to reveal physical-layer attacks, which does not meet the reliability requirements of QRNGs in high-security applications.

\section{Conclusion}
\label{sec:conclusion}

This work highlights a critical consideration for the security validation of Quantum Random Number Generators, particularly within the common paradigm that pairs powerful randomness extraction with final-stage statistical testing. Globally, standards such as the NIST SP 800-22 suite from the United States, Germany's AIS-31\cite{killmann2011proposal} which is widely used for cryptographic applications within the European Union\cite{petura2016survey}, and China's GM/T 0005–2021\cite{Luo2022GMT0005} form the bedrock of the certification process. Our study does not question the validity of these statistical tests themselves, but rather demonstrates a scenario where their effectiveness as a security sentinel is compromised. We show that standard post-processing algorithms, in fulfilling their mathematical function of normalizing statistical distributions, can inadvertently conceal the effects of a successful physical-layer attack, potentially creating a false sense of security.

Our research substantiates this claim through a practical power supply ripple attack. By injecting a periodic disturbance into the ASE-based QRNG's drive circuitry, the physical entropy source was significantly compromised. The resultant raw data exhibited a high correlation with the attack signal and consequently failed all tests within the NIST suite. However, following the application of a conventional Toeplitz-hash randomness extractor, the same compromised data successfully passed all 15 NIST tests.

This outcome reveals a significant security challenge: the combination of a powerful randomness extractor and a final-stage statistical test suite can validate data from a completely compromised device. It establishes that a QRNG can satisfy standard certification criteria while its core quantum process is, in fact, subject to external control. The fact that our heavily manipulated data passed all tests implies that even if an attacker achieved complete control—making the raw stream fully predictable—this standard validation pipeline would still certify the output as secure. The implications for any application relying on such a device are catastrophic. For a cryptographic system, if these compromised numbers are used to generate secret keys, nonces, or initialization vectors, its security is effectively nullified. An attacker aware of the physical manipulation could predict the random output, fatally undermining the protocol's security foundation.

Consequently, our findings suggest that reliance solely on the statistical properties of the final output is an insufficient security methodology against physical-layer threats. As QRNGs become integrated into increasingly sensitive applications, it is imperative that future security frameworks evolve. Such frameworks must complement rigorous statistical validation with the real-time integrity monitoring of the physical entropy source itself, thereby safeguarding against these concealed vulnerabilities.



\bibliography{apssamp}

\end{document}